\documentclass[
    ,final            
  ]
  {aipproc}

\layoutstyle{8x11single}
\usepackage{multirow}

\begin{document}

\title{Low-mass stars within dense dark matter halos}


\author{Jordi Casanellas}{
  address={Centro Multidisciplinar de Astrof\'{\i}sica, Instituto Superior T\'ecnico, Av. Rovisco Pais, 1049-001 Lisboa, Portugal}
}
\author{Il\'\i dio Lopes}{
  address={Departamento de F\'\i sica,
Universidade de \'Evora, Col\'egio Ant\'onio Luis
Verney, 7002-554 \'Evora - Portugal}
  ,altaddress={Centro Multidisciplinar de Astrof\'{\i}sica, Instituto Superior T\'ecnico, Av. Rovisco Pais, 1049-001 Lisboa, Portugal}
}

\begin{abstract}
We studied the formation and evolution of low-mass stars within halos with high concentration of dark matter (DM) particles, using a highly sophisticated expression to calculate the rate at which DM particles are captured inside the star. For very high DM densities in the host halo ($\rho_{\chi}>10^{10}\;$GeV$\;$cm$^{-3}$ for a 1 M$_{\odot}$ star), we found that young stars stop sooner their gravitational collapse in the pre-Main Sequence phase, reaching states of equilibrium in which DM annihilation is their only source of energy. The lower effective temperature of these stars, which depends on the properties of the DM particles and DM halo, may be used as an alternative method to investigate the nature of DM.\\
\end{abstract}

\maketitle

\section{Introduction}

Cosmological observations at different scales suggest that $\sim96\%$ of the Universe is composed of invisible matter and energy. From that ammount, $\sim23\%$ is known as Dark Matter (DM), which is believed to be partially responsible for the formation of the first structures of the Universe. If the DM component is composed by particles, they are very likely WIMPs (Weakly Interacting Massive Particles), given that they must be stable, neutral, and massive. WIMPs have sizeable scattering cross sections with baryons, which upper limits are set by direct detection experiments \cite{art-XENON10_SD2008, art-CDMSII_SI2009}, and their annihilation cross section is known to be of the order $<\sigma_a v>=3\cdot10^{-26}\;$cm$^3\;$s$^{-1}$ \cite{rev-BertoneHS2005}. These two properties make DM particles accumulate inside stars and annihilate among themselves, providing a new source of energy for the star \cite{art-PressSpergel1985,art-SalatiSilk1989}.

Recently, many authors have been studying the influence of self-annihilating DM particles in the first generation of stars \cite{let-Spolyaretal2008,art-Ioccoetal2008,art-FreeseBSG08,art-Taosoetal2008,art-YoonIoccoAkiyama2008,art-Natarajan2009,art-RipamontiIoccoetal2009}, in compact objects \cite{art-MoskalenkoWai2007,art-BertoneFairbairn2008,art-Kouvaris2008}, in the Sun \cite{let-LopesSilk2002, art-LopesBS2002, art-LopesSH2002, art-Bottinoetal2002}, and in low-mass stars of the Main Sequence (MS) \cite{art-Scottetal2007,art-Fairbairnetal2008,art-Scottetal2009}. In this proceedings we will review the new stellar evolution scenarios that low-mass stars follow when formed in dense halos of DM, which are described in detail in our work \cite{art-CasanellasLopes2009}, but this time we will use an upgraded calculation of the process of capture of DM particles by the star which, compared with our previous results, leads to stronger effects in the forming star in the pre-MS phase. Nevertheless, the overall results and conclusions are similar.

\subsection{Capture and annihilation of DM particles}

The evolution of stars in DM halos is regulated by the capture of DM particles in the stellar interior. DM particles in the halo may scatter with the nuclei of elements inside the star. If in these collisions the DM particles lose enough energy, they will get gravitationally trapped in the core of the star. To compute the number of particles that are captured per second by the star, we adapted part of the publicly available \texttt{DarkSUSY} code \cite{art-GondoloEdsjoDarkSusy2004}, which calculates the capture rate $C_{\chi}$ from the original expressions of A.Gould \cite{art-Gould1987}:
\begin{equation}
\label{cap}
    C_{\chi}(t) = \int^{R_\star}_0 4\pi r^2\int^\infty_0 \frac{f_{v_{\star}}(u)}{u}w\Omega_v^-(w)\,\mathrm{d}u\,\mathrm{d}r\;,
\end{equation}
where $\Omega_v^-(w)$ is the probability of a DM particle with a velocity $w$ to have, after the collision with the nucleus of an element i, a velocity $v$ lower than the escape velocity of the star $v_{esc,r}$ at the radius of the collision, 
\begin{equation}
 \Omega_v^-(w) = \sum_i \frac{\sigma_i n_i(r,t)}{w}\Big(v_{esc,r}^2-\frac{\mu_{-,i}^2}{\mu_i}u^2\Big)\theta\Big(v_{esc,r}^2-\frac{\mu_{-,i}^2}{\mu_i}u^2\Big),
\end{equation}
\begin{equation}
    \mu\equiv\frac{m_\chi}{m_\mathrm{n}}, \quad\mu_\pm\equiv\frac{\mu\pm 1}{2}\;.
\end{equation}
In our computations we calculated the capture rate for 16 elements (H, $^4$He, $^3$He, $^{12}$C, $^{14}$N, $^{16}$O, Ne, Na, Mg, Al, Si, S, Ar, Ca, Fe and Ni), which abundances are followed by our code. The initial chemical abundances and initial metallicity of the stars were assumed to be as the solar ones \cite{art-AsplundGrevesseSauval2005}. All elements except hydrogen have negligible contributions to the total capture rate. This is because H is the only element with spin dependent (SD) interaction with the DM particles, and the experimental scattering cross section limits for this type of interaction is less stringent than for the spin independent (SI) one.

The calculation of the capture rate implemented in the present work takes into account that the DM particles have a velocity distribution $f_0(u)$, which we assumed to be a Maxwell-Boltzmann distribution with a dispersion velocity $\bar{v_{\chi}}$. This $C_{\chi}$ expression also permits the consideration of different velocities of the star $v_\star$, leading to a distribution of the velocities of the DM particles seen by the star $f_{v_{\star}}(u)$ equal to:
\begin{equation}
    f_{v_{\star}}(u) = f_0(u) \exp\Big(-\frac{3v_\star^2}{2\bar{v_{\chi}}^2}\Big) \frac{\sinh(3uv_\star/\bar{v_{\chi}}^2)}{3uv_\star/\bar{v_{\chi}}^2}
\label{eq-fvs}
\end{equation}
\begin{equation}
    f_0(u) = \frac{\rho_\chi}{m_\chi} \frac{4}{\sqrt{\pi}} \Big(\frac{3}{2}\Big)^{3/2} \frac{u^2}{\bar{v_{\chi}}^3} \exp\Big(-\frac{3u^2}{2\bar{v_{\chi}}^2}\Big).
\label{eq-fv0}
\end{equation}

In table \ref{tab-Cx} are shown the capture rates obtained for stars at different stages of their evolution, considering different stellar masses $M_{\star}$ and velocities $v_{\star}$, and different values for the mass of the DM particles $m_{\chi}$. Higher capture rates were obtained during the MS, for stars with higher $M_{\star}$ and lower $v_{\star}$, and for DM particles with lower $m_{\chi}$.

\begin{table}
\hspace{2.5cm}
\begin{tabular}{c c|c}
\hline
& & \tablehead{1}{r}{b}{$\mathbf{C_{\chi}}$ (s$^{-1}$)} \\
\hline
\multirow{2}{*}{\textbf{Stage}} & Pre-MS\tablenote{\hspace{2.5cm}$^*$(when $L=10\;$L$_{\odot}$, $T_{eff}=4610\;$K)} & $1\times10^{31}$ \\
& ZAMS & $9\times10^{32}$ \\
&&\\
\multirow{2}{*}{$\mathbf{M_{\star}}$ (M$_{\odot}$)} & 0.7 & $4\times10^{32}$ \\
& 3.0 & $8\times10^{33}$ \\
&&\\
\multirow{2}{*}{$\mathbf{v_{\star}}$ (km s$^{-1}$)} & 50  & $2\times10^{33}$ \\
& 350 & $3\times10^{32}$ \\
&&\\
\multirow{2}{*}{$\mathbf{m_{\chi}}$ (GeV)} & 10 & $3\times10^{34}$ \\
& 1000 & $1\times10^{31}$ \\
\hline
\end{tabular}
\hspace{2.5cm}.
\caption{Capture rates obtained for stars at different stages of their evolution, for different stellar masses, stellar velocities, and mass of the DM particles. If not stated otherwise, the capture rates are those when stars are in the Zero Age Main Sequence (ZAMS), with $M_{\star}=1\;$M$_{\odot}$, $v_\star=220\;$km s$^{-1}$, and $m_{\chi}=100\;$GeV. In all cases, the stars evolved in a halo of DM particles with $\rho_{\chi}=10^8\;$GeV$\;$cm$^{-3}$, ${v_{\chi}}=270\;$km s$^{-1}$, and $\sigma_{\chi,SD}=10^{-38}\;$cm$^2$.}
\label{tab-Cx}
\end{table}

Once they are captured, DM particles accumulate in the center of the star, in a region with a typical radius $r_\chi=\sqrt{3\kappa T_c/2\pi G \rho_c m_\chi}$. There, WIMPs annihilate among themselves, thus providing a new source of energy that contributes to the total luminosity of the star with:
\begin{equation}
L_\chi= f_\chi\;C_\chi\;m_\chi,
\end{equation}
where $f_\chi=2/3$ is a factor to take into account that a third of the products of DM annihilation will escape out of the star in the form of neutrinos.

\section{Stellar evolution scenarios within dense dark matter halos}
\begin{figure}[!t]
\includegraphics[]{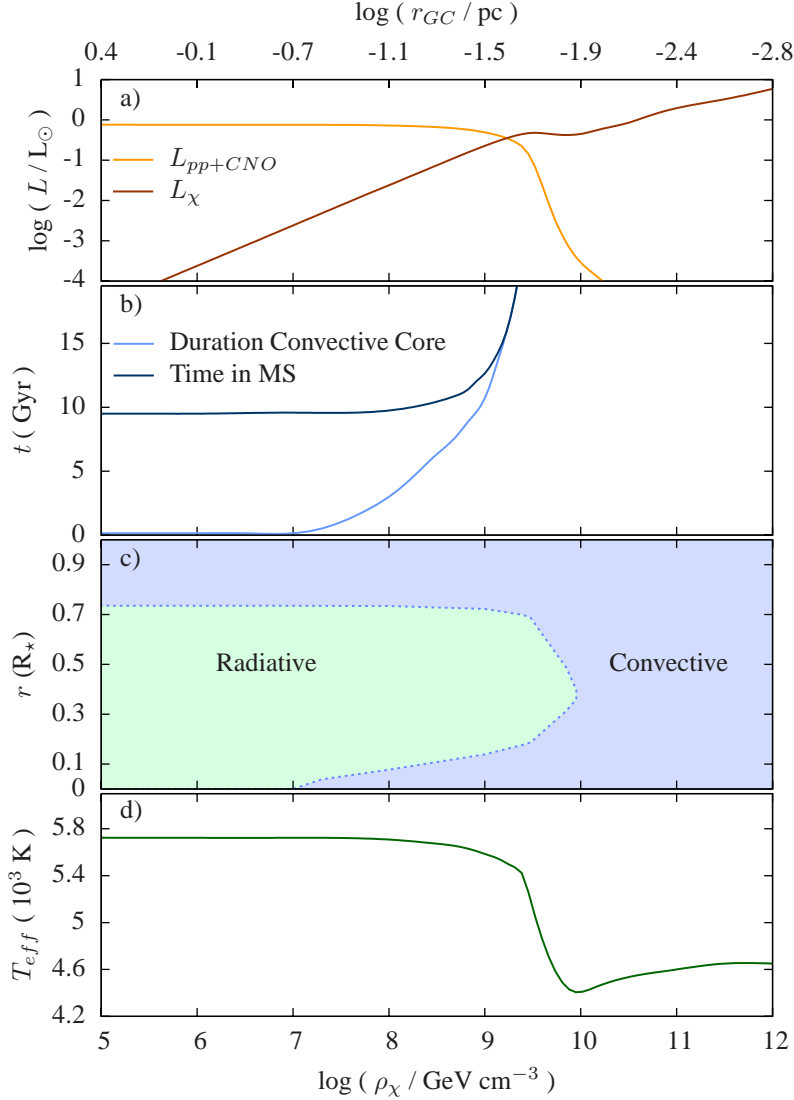}
\caption{Properties of stars of 1 M$_{\odot}$ that evolve in halos with DM densities from 10$^5$ to 10$^{12}$ GeV cm$^{-3}$: a) luminosity from the thermonuclear reactions and from DM annihilation; b) time spent by the star in the Main Sequence (MS) (from $\varepsilon_{grav}<1\%\,\varepsilon_{T}$ to $X_{c} < 0.001$) and time until the convective core disappears; c) location of the convective and radiative zones inside the star; d) effective temperature of the star. The characteristics in Figures a), c) and d) are plotted at such an age that all the stars are already in energy equilibrium, i.e.: in the beginning of the MS for $\rho_{\chi}<3\times10^9\;$GeV$\;$cm$^{-3}$, and in the stationary states powered only by the energy from DM annihilation for higher DM densities.}
\label{fig-multi}
\end{figure}
\begin{figure}
  \includegraphics[]{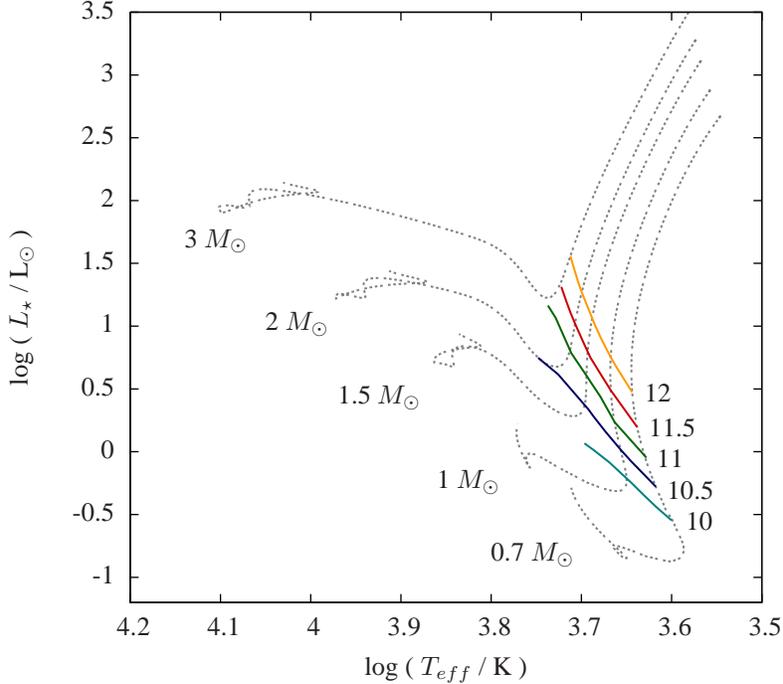}
  \caption{Stationary states reached by stars with masses from 0.7 to 3$\;$M$_{\odot}$ when the energy from DM annihilation compensates the gravitational energy during the collapse. These equilibrium positions, where stars will remain for an indefinite time, are plotted for different dark matter halo densities, indicated in units of $\log(\rho_{\chi}/$GeV$\;$cm$^{-3}$) at the side of each line. The grey lines are the classical evolutionary paths.}
\label{fig-statHR}
\end{figure}

The way DM influences the star depends on the amount of energy released by DM annihilation. In this section, this quantity depends mainly on the DM density on the host halo $\rho_{\chi}$, given that the other parameters are fixed to the following fiducial values: $v_\star=220\;$km s$^{-1}$, $\bar{v_{\chi}}=270\;$km s$^{-1}$, $\sigma_{\chi,SD}=10^{-38}\;$cm$^2$, $\sigma_{\chi,SI}=10^{-44}\;$cm$^2$, and $m_{\chi}=100\;$GeV.

For "low" DM densities ($\rho_{\chi}<3\times10^9\;$GeV$\;$cm$^{-3}$ for a star of 1 M$_{\odot}$), the energy from DM annihilation is a complementary source of energy for the star (see Figure \ref{fig-multi}.a)). Their main impacts on stellar evolution are: i) DM burning contributes, together with the thermonuclear energy, to compensate the gravitational collapse of the protostar, which will reach the hydrostatic equilibrium at a lower central temperature than in the classical scenario, leading to a lower rate of hydrogen burning and, therefore, an extension of the lifetime of the star in the MS (see Figure \ref{fig-multi}.b)); ii) as the energy from DM annihilation is produced in a very concentrated region ($r_{\chi}\sim0.03\;R_{\star}$), the gradient of temperature in the core is very high, requiring the convective core to remain for a longer time than in the classical scenario in order to evacuate the extra input of energy in a more efficient way (see Figure \ref{fig-multi}.b)).

For high DM densities, the energy from the annihilation of captured DM particles compensates the gravitational collapse of the protostar when the central temperature is still below the threshold for thermonuclear reactions. In this new scenario, the star remains in an indefinite state of equilibrium with DM burning as its only source of energy. In Figure \ref{fig-statHR}, we show the equilibrium positions in the Hertzsprung-Russell diagram reached by stars with masses between 0.7 and 3 M$_{\odot}$ that evolved in halos with different DM densities. These results show that stars "freeze" during their pre-MS track sooner than in our previous study \cite{art-CasanellasLopes2009}, which was carried out with a more basic capture rate calculation. Regarding the energy transport mechanisms inside the star, convection is the only one capable of transporting the huge amount of energy through the whole star if it evolves in halos with very high DM densities. In the case of a star of 1 M$_{\odot}$, the star is fully convective for $\rho_{\chi}>10^{10}\;$GeV$\;$cm$^{-3}$. For DM densities between 10$^{9}$ and $10^{10}\;$GeV$\;$cm$^{-3}$, the star will remain with a convective core for the rest of its life (see Figure \ref{fig-multi}.b) and c)). One of the direct consequences of the changes in the extension of the convective zones is a reduction on the effective temperature of the star $T_{eff}$. For a one solar mass star, the drop on $T_{eff}$ is of the order of $\sim1000\;$K (see Figure \ref{fig-multi}.c)).

\section{Conclusions}

The search for the unusual stars described in this manuscript appears as a promising alternative method to test and constrain the properties of DM candidates. In the higher horizontal axe of Figure \ref{fig-multi}, are shown the expected distances toward the center of our galaxy where stars with such unusual properties may be found, following the DM density profile of Bertone \& Merrit \cite{art-BertonteMerritt2005}. These distances are shown to provide an indication of where to find these stars within the inner parsec of our galaxy. Many parameters with high uncertainties influence this prediction: the velocity distribution of the DM particles, their scattering cross sections with baryons and the DM density profile, among others. The orbit followed by these stars also plays an important role: the changes on the stellar velocity with respect to the halo can boost or diminish the capture rate (see table \ref{tab-Cx} and reference \cite{art-Scottetal2009}), varying the contribution of DM burning to the total luminosity. We expect that with the future improvements on the observation of stars near the galactic center we will be able to give constraints to the DM particle models, providing an alternative approach to reveal the nature of Dark Matter.

\begin{theacknowledgments}
We are grateful to Fabio Iocco, for his helpful comments, and to the authors of the \texttt{DarkSUSY} code, for making the code publicly available. This work was supported by grants from "Funda\c c\~ao para a Ci\^encia
e Tecnologia" (SFRH/BD/44321/2008) and "Funda\c c\~ao Calouste Gulbenkian".
\end{theacknowledgments}



\bibliographystyle{aipproc}   

\bibliography{DM}

\begin{thebibliography}{27}
\expandafter\ifx\csname natexlab\endcsname\relax\def\natexlab#1{#1}\fi
\providecommand{\enquote}[1]{``#1''}
\expandafter\ifx\csname url\endcsname\relax
  \def\url#1{\texttt{#1}}\fi
\expandafter\ifx\csname urlprefix\endcsname\relax\def\urlprefix{URL }\fi
\providecommand{\eprint}[2][]{\url{#2}}

\bibitem[Angle et~al.(2008)]{art-XENON10_SD2008}
J.~Angle, et~al., \emph{Phys. Rev. Lett.} \textbf{101}, 091301 (2008),
  \eprint{arXiv:0805.2939}.

\bibitem[Ahmed et~al.(2009)]{art-CDMSII_SI2009}
Z.~Ahmed, et~al., \emph{Phys. Rev. Lett.} \textbf{102}, 011301--+ (2009),
  \eprint{arXiv:0802.3530}.

\bibitem[{Bertone} et~al.(2005)]{rev-BertoneHS2005}
G.~{Bertone}, D.~{Hooper}, and J.~{Silk}, \emph{Phys. Rep.} \textbf{405},
  279--390 (2005), \eprint{arXiv:hep-ph/0404175}.

\bibitem[{Press} and {Spergel}(1985)]{art-PressSpergel1985}
W.~H. {Press}, and D.~N. {Spergel}, \emph{ApJ} \textbf{296}, 679--684 (1985).

\bibitem[{Salati} and {Silk}(1989)]{art-SalatiSilk1989}
P.~{Salati}, and J.~{Silk}, \emph{ApJ} \textbf{338}, 24--31 (1989).

\bibitem[{Spolyar} et~al.(2008)]{let-Spolyaretal2008}
D.~{Spolyar}, K.~{Freese}, and P.~{Gondolo}, \emph{Physical Review Letters}
  \textbf{100}, 051101--+ (2008), \eprint{arXiv:0705.0521}.

\bibitem[{Iocco} et~al.(2008)]{art-Ioccoetal2008}
F.~{Iocco}, A.~{Bressan}, E.~{Ripamonti}, R.~{Schneider}, A.~{Ferrara}, and
  P.~{Marigo}, \emph{MNRAS} \textbf{390}, 1655--1669 (2008),
  \eprint{arXiv:0805.4016}.

\bibitem[{Freese} et~al.(2008)]{art-FreeseBSG08}
K.~{Freese}, P.~{Bodenheimer}, D.~{Spolyar}, and P.~{Gondolo}, \emph{ApJ}
  \textbf{685}, L101--L104 (2008), \eprint{arXiv:0806.0617}.

\bibitem[{Taoso} et~al.(2008)]{art-Taosoetal2008}
M.~{Taoso}, G.~{Bertone}, G.~{Meynet}, and S.~{Ekstr{\"o}m}, \emph{Phys. Rev.
  D} \textbf{78}, 123510--+ (2008), \eprint{arXiv:0806.2681}.

\bibitem[{Yoon} et~al.(2008)]{art-YoonIoccoAkiyama2008}
S.-C. {Yoon}, F.~{Iocco}, and S.~{Akiyama}, \emph{ApJ} \textbf{688}, L1--L4
  (2008), \eprint{arXiv:0806.2662}.

\bibitem[{Natarajan} et~al.(2009)]{art-Natarajan2009}
A.~{Natarajan}, J.~C. {Tan}, and B.~W. {O'Shea}, \emph{ApJ} \textbf{692},
  574--583 (2009), \eprint{arXiv:0807.3769}.

\bibitem[{Ripamonti} et~al.(2009)]{art-RipamontiIoccoetal2009}
E.~{Ripamonti}, F.~{Iocco}, A.~{Bressan}, R.~{Schneider}, A.~{Ferrara}, and
  P.~{Marigo}  (2009), \eprint{arXiv:0903.0346}.

\bibitem[{Moskalenko} and {Wai}(2007)]{art-MoskalenkoWai2007}
I.~V. {Moskalenko}, and L.~L. {Wai}, \emph{ApJ} \textbf{659}, L29--L32 (2007),
  \eprint{arXiv:astro-ph/0702654}.

\bibitem[{Bertone} and {Fairbairn}(2008)]{art-BertoneFairbairn2008}
G.~{Bertone}, and M.~{Fairbairn}, \emph{Phys. Rev. D} \textbf{77}, 043515--+
  (2008), \eprint{arXiv:0709.1485}.

\bibitem[{Kouvaris}(2008)]{art-Kouvaris2008}
C.~{Kouvaris}, \emph{Phys. Rev. D} \textbf{77}, 023006--+ (2008),
  \eprint{arXiv:0708.2362}.

\bibitem[{Lopes} and {Silk}(2002)]{let-LopesSilk2002}
I.~P. {Lopes}, and J.~{Silk}, \emph{Physical Review Letters} \textbf{88},
  151303--+ (2002), \eprint{arXiv:astro-ph/0112390}.

\bibitem[{Lopes} et~al.(2002{\natexlab{a}})]{art-LopesBS2002}
I.~P. {Lopes}, G.~{Bertone}, and J.~{Silk}, \emph{MNRAS} \textbf{337},
  1179--1184 (2002{\natexlab{a}}), \eprint{arXiv:astro-ph/0205066}.

\bibitem[{Lopes} et~al.(2002{\natexlab{b}})]{art-LopesSH2002}
I.~P. {Lopes}, J.~{Silk}, and S.~H. {Hansen}, \emph{MNRAS} \textbf{331},
  361--368 (2002{\natexlab{b}}), \eprint{arXiv:astro-ph/0111530}.

\bibitem[{Bottino} et~al.(2002)]{art-Bottinoetal2002}
A.~{Bottino}, G.~{Fiorentini}, N.~{Fornengo}, B.~{Ricci}, S.~{Scopel}, and
  F.~L. {Villante}, \emph{Phys. Rev. D} \textbf{66}, 053005--+ (2002),
  \eprint{arXiv:hep-ph/0206211}.

\bibitem[{Scott} et~al.(2007)]{art-Scottetal2007}
P.~{Scott}, J.~{Edsj{\"o}}, and M.~{Fairbairn}  (2007),
  \eprint{arXiv:0711.0991}.

\bibitem[{Fairbairn} et~al.(2008)]{art-Fairbairnetal2008}
M.~{Fairbairn}, P.~{Scott}, and J.~{Edsj{\"o}}, \emph{Phys. Rev. D}
  \textbf{77}, 047301--+ (2008), \eprint{arXiv:0710.3396}.

\bibitem[{Scott} et~al.(2009)]{art-Scottetal2009}
P.~{Scott}, M.~{Fairbairn}, and J.~{Edsj{\"o}}, \emph{MNRAS} \textbf{394},
  82--104 (2009), \eprint{arXiv:0809.1871}.

\bibitem[{Casanellas} and {Lopes}(2009)]{art-CasanellasLopes2009}
J.~{Casanellas}, and I.~{Lopes}, \emph{ApJ} \textbf{705}, 135--143 (2009),
  \eprint{arXiv:0909.1971}.

\bibitem[{Gondolo} et~al.(2004)]{art-GondoloEdsjoDarkSusy2004}
P.~{Gondolo}, J.~{Edsj{\"o}}, P.~{Ullio}, L.~{Bergstr{\"o}m}, M.~{Schelke}, and
  E.~A. {Baltz}, \emph{Journal of Cosmology and Astro-Particle Physics}
  \textbf{7}, 8--+ (2004), \eprint{arXiv:astro-ph/0406204}.

\bibitem[{Gould}(1987)]{art-Gould1987}
A.~{Gould}, \emph{ApJ} \textbf{321}, 571--585 (1987).

\bibitem[{Asplund} et~al.(2005)]{art-AsplundGrevesseSauval2005}
M.~{Asplund}, N.~{Grevesse}, and A.~J. {Sauval}, \enquote{{The Solar Chemical
  Composition},} in \emph{Cosmic Abundances as Records of Stellar Evolution and
  Nucleosynthesis}, edited by T.~G. {Barnes}, III, and F.~N. {Bash}, 2005, vol.
  336 of \emph{Astronomical Society of the Pacific Conference Series}, pp.
  25--+.

\bibitem[{Bertone} and {Merritt}(2005)]{art-BertonteMerritt2005}
G.~{Bertone}, and D.~{Merritt}, \emph{Modern Physics Letters A} \textbf{20},
  1021--1036 (2005), \eprint{arXiv:astro-ph/0504422}.

\end{thebibliography}

\IfFileExists{\jobname.bbl}{}
 {\typeout{}
  \typeout{******************************************}
  \typeout{** Please run "bibtex \jobname" to optain}
  \typeout{** the bibliography and then re-run LaTeX}
  \typeout{** twice to fix the references!}
  \typeout{******************************************}
  \typeout{}
 }

\end{document}